\magnification=\magstep1 
\font\bigbfont=cmbx10 scaled\magstep1
\font\bigifont=cmti10 scaled\magstep1
\font\bigrfont=cmr10 scaled\magstep1

\vsize = 23.5 truecm
\hsize = 15.5 truecm
\hoffset = .2truein
\baselineskip = 14 truept
\edef\cite{\the\catcode`@}%
\catcode`@ = 11
\let\@oldatcatcode = \cite
\edef\onlinecite{\the\catcode`@}%
\let\@oldatcatcode = \onlinecite
\chardef\@letter = 11
\chardef\@other = 12
%
%
%
%
\def\@gobble#1{}%
%
%
%
%
\def\@tokstostring#1{\@ttsA#1 \@ttsmarkA}%
%
%
\def\@ttsA#1 #2\@ttsmarkA{%
   \@ifempty{#1}\else
      \@ttsB #1\@ttsmarkB
      \@ifempty{#2}\else
         \@spacesub\@ttsA#2\@ttsmarkA
      \fi
   \fi
}%
%
%
\def\@ttsB#1{%
   \ifx #1\@ttsmarkB\else
      \string #1%
      \expandafter\@ttsB
   \fi
}%
%
%
\def\@ttsmarkB{\@ttsmarkB}
%
%
\def\@spacesub{+}%
%
%
%
\def\@ifempty#1{\@@ifempty #1\@emptymarkA\@emptymarkB}%
\def\@@ifempty#1#2\@emptymarkB{\ifx #1\@emptymarkA}%
\def\@emptymarkA{\@emptymarkA}
%
%
%
\def\@nnil{\@nil}%
\def\@fornoop#1\@@#2#3{}%
\def\@for#1:=#2\do#3{%
   \edef\@fortmp{#2}%
   \ifx\@fortmp\empty \else
      \expandafter\@forloop#2,\@nil,\@nil\@@#1{#3}%
   \fi
}%
\def\@forloop#1,#2,#3\@@#4#5{\def#4{#1}\ifx #4\@nnil \else
       #5\def#4{#2}\ifx #4\@nnil \else#5\@iforloop #3\@@#4{#5}\fi\fi
}%
\def\@iforloop#1,#2\@@#3#4{\def#3{#1}\ifx #3\@nnil
       \let\@nextwhile=\@fornoop \else
      #4\relax\let\@nextwhile=\@iforloop\fi\@nextwhile#2\@@#3{#4}%
}%
%
%
%
\newif\if@fileexists
\def\@testfileexistence#1{%
   \begingroup
      \immediate\openin0 = \jobname.#1 
      \ifeof 0
         \global\@fileexistsfalse
      \else
         \global\@fileexiststrue
      \fi
      \immediate\closein0
   \endgroup
}%
%
%
%
\def\bibliographystyle#1{\write\@auxfile{\string\bibstyle{#1}}}%
\let\bibstyle = \@gobble
%
%
\def\bibliography#1{%
   \write\@auxfile{\string\bibdata{#1}}%
   \@testfileexistence{bbl}%
   \if@fileexists
      \@readbblfile
   \fi
}%
\let\bibdata = \@gobble
%
%
%
\def\nocite#1{\write\@auxfile{\string\citation{#1}}}%
\newif\if@notfirstcitation
%
%
\def\onlinecite{\futurelet\next\@onlinecite}%
\def\@onlinecite{%
   \begingroup \if [\next
      \aftergroup\@citewithnote
   \else
      \global\let\@citenote = \empty
      \aftergroup\@@onlinecite
   \fi \endgroup
}%
\def\cite{\futurelet\next\@cite}%
\def\@cite{%
   \begingroup \if [\next
      \aftergroup\@citewithnote
   \else
      \global\let\@citenote = \empty
      \aftergroup\@@cite
   \fi \endgroup
}%
%
%
\def\@citewithnote[#1]{%
   \def\@citenote{#1}%
   \@@cite
}%
%
%
\def\@@cite#1{%
   \nocite{#1}%
   %
   \printcitestart
   \@notfirstcitationfalse
   \@for \@citation :=#1\do
   {%
      \expandafter\@onecitation\@citation\@@
   }%
   \ifx\empty\@citenote\else
      \printcitenote{\@citenote}%
   \fi
   \printcitefinish
}%
\def\@onecitation#1\@@{%
   \if@notfirstcitation
      \printbetweencitations
   \fi
   \expandafter \ifx \csname\@citelabel{#1}\endcsname \relax
      \if@citewarning
         \message{Undefined citation `#1'.}%
      \fi
      \expandafter\gdef\csname\@citelabel{#1}\endcsname{%
         \nobreak\hskip0pt#1\nobreak\hskip0pt}%
   \fi
   \csname\@citelabel{#1}\endcsname
   \@notfirstcitationtrue
}%
\def\@@onlinecite#1{%
   \nocite{#1}%
   %
   \printonlinecitestart
   \@notfirstcitationfalse
   \@for \@citation :=#1\do
   {%
      \expandafter\@onecitation\@citation\@@
   }%
   \ifx\empty\@citenote\else
      \printcitenote{\@citenote}%
   \fi
   \printonlinecitefinish
}%
\def\@oneonelinecitation#1\@@{%
   \if@notfirstcitation
      \printbetweenonlinecitations
   \fi
   \expandafter \ifx \csname\@citelabel{#1}\endcsname \relax
      \if@citewarning
         \message{Undefined citation `#1'.}%
      \fi
      \expandafter\gdef\csname\@citelabel{#1}\endcsname{%
         \nobreak\hskip0pt#1\nobreak\hskip0pt}%
   \fi
   \csname\@citelabel{#1}\endcsname
   \@notfirstcitationtrue
}%
%
%
\def\@citelabel#1{\@tokstostring{b@#1}}%
%
%
\def\@citedef#1{\@resetnumerals\@finishcitedef{#1}}%
\def\@finishcitedef#1#2{\expandafter\gdef\csname\@citelabel{#1}\endcsname{#2}}%
\def\@resetnumerals{%
   \catcode`0 = \@other \catcode`1 = \@other \catcode`2 = \@other
   \catcode`3 = \@other \catcode`4 = \@other \catcode`5 = \@other
   \catcode`6 = \@other \catcode`7 = \@other \catcode`8 = \@other
   \catcode`9 = \@other
}%
%
%
%
\def\@readbblfile{%
   \begingroup
      \def\begin##1##2{%
         \setbox0 = \hbox{\biblabelcontents{##2}}%
         \biblabelwidth = \wd0
      }%
      \def\end##1{}
      %
      %
      \@itemnum = 0
      \def\bibitem{\futurelet\next\@bibitem}%
      \def\@bibitem{%
         \begingroup \if [\next
            \aftergroup\@alphabibitem
         \else
            \aftergroup\@numberedbibitem
         \fi \endgroup
      }%
      \def\@alphabibitem[##1]##2{%
         \expandafter \xdef\csname\@citelabel{##2}\endcsname {##1}%
         \@finishbibitem{##2}%
      }%
      \def\@numberedbibitem##1{%
         \advance\@itemnum by 1
         \expandafter \xdef\csname\@citelabel{##1}\endcsname{\number\@itemnum}%
         \@finishbibitem{##1}%
      }%
      \def\@finishbibitem##1{%
         \biblabelprint{\csname\@citelabel{##1}\endcsname}%
         \write\@auxfile
            {\string\@citedef{##1}{\csname\@citelabel{##1}\endcsname}}%
         \ignorespaces
      }%
      %
      %
      \let\em = \bblem
      \let\newblock = \bblnewblock
      \frenchspacing
      \clubpenalty = 4000 \widowpenalty = 4000
      \tolerance = 10000 \hfuzz = .5pt
      \parskip = 1.5ex plus .5ex minus .5ex
      \everypar = {\hangindent = \biblabelwidth \advance\hangindent by .5em}%
      \bblrm
      \bblhook
      \input \jobname.bbl
   \endgroup
}%
%
%
\newcount\@itemnum
%
%
\newdimen\biblabelwidth
%
%
%
%
\def\biblabelprint#1{%
   \noindent\hbox to \biblabelwidth{\biblabelcontents{#1}\hfil}\enspace}%
%
%
\def\biblabelcontents#1{\bblrm [#1]}%
%
%
\def\bblrm{\rm}%
%
%
\def\bblem{\it}%
%
%
\def\bblnewblock{\hskip .11em plus .33em minus .07em}%
%
%
\let\bblhook = \empty
%
%
%
\def\printonlinecitestart{}
\def\printonlinecitefinish{}
\def\printbetweenonlinecitations{, }
\def\printcitenote#1{, #1}
%
     \def\printcitestart{$^\bgroup}
     \def\printcitefinish{\egroup$}
     \def\printbetweencitations{,}
%
%
\let\citation = \@gobble
%
%
%
\newwrite\@auxfile
\newif\if@auxfileopened
\def\@openauxfile{%
   \if@auxfileopened\else
      \@auxfileopenedtrue
      \immediate\openout\@auxfile = \jobname.aux
   \fi
}%
%
%
%
\def\@readauxfile{%
   \@testfileexistence{aux}%
   \if@fileexists
      \begingroup
         \@setletters
         \input \jobname.aux
      \endgroup
   \else
      \message{\@undefinedmessage}%
      \@citewarningfalse
   \fi
   \@openauxfile
}%
%
%
\def\@setletters{%
   \catcode`_ = \@letter
   \catcode`+ = \@letter \catcode`- = \@letter
   \catcode`@ = \@letter
   \catcode`0 = \@letter \catcode`1 = \@letter \catcode`2 = \@letter
   \catcode`3 = \@letter \catcode`4 = \@letter \catcode`5 = \@letter
   \catcode`6 = \@letter \catcode`7 = \@letter \catcode`8 = \@letter
   \catcode`9 = \@letter
   \catcode`( = \@letter \catcode`) = \@letter
   \@setmoreletters
}%
%
%
\ifx\@setmoreletters\@undefined
   \let\@setmoreletters = \empty
\fi
%
%
\ifx\@undefinedmessage\@undefined
   \def\@undefinedmessage{No .aux file; I won't give you warnings about
                          undefined citations.}%
\fi
%
%
\newif\if@citewarning  \@citewarningtrue
%
%
%
\newcount\@numparams
%
%
\def\newcommand#1{%
   \def\@commandname{#1}%
   \futurelet\@next\@continuenewcommand
}%
%
%
\def\@continuenewcommand{\begingroup
   \if [\@next
      \aftergroup\@newcommandwithargs
   \else
      \global\@numparams = 0
      \aftergroup\@newcommand
   \fi
\endgroup}%
\def\@newcommandwithargs[#1]{%
   \global\@numparams = #1
   \@newcommand
}%
%
%
\def\@newcommand#1{%
   \def\@startdef{\expandafter\edef\@commandname}%
   \ifnum\@numparams=0
      \let\@paramdef = \empty
   \else
      \ifnum\@numparams>9
         \errmessage{\the\@numparams\space is too many parameters}%
      \else
         \ifnum\@numparams<0
            \errmessage{\the\@numparams\space is too few parameters}%
         \else
            \edef\@paramdef{%
               %
               \ifcase\@numparams
                  \empty  No arguments.
               \or ####1%
               \or ####1####2%
               \or ####1####2####3%
               \or ####1####2####3####4%
               \or ####1####2####3####4####5%
               \or ####1####2####3####4####5####6%
               \or ####1####2####3####4####5####6####7%
               \or ####1####2####3####4####5####6####7####8%
               \or ####1####2####3####4####5####6####7####8####9%
               \fi
            }%
         \fi
      \fi
   \fi
   \expandafter\@startdef\@paramdef{#1}%
}%
%
%
%
\@readauxfile
%
%
\catcode`@ = \@oldatcatcode

\input epsf
\nopagenumbers
\def\bibitem#1#2{\item{[#2]}}
%
%
\topinsert
\vskip 3.2 truecm
\endinsert
%
%
\centerline{\bigbfont IMPURITY DYNAMICS IN A BOSE CONDENSATE}
%
%
\vskip 20 truept
\centerline{\bigifont Siu A. Chin and  Harald A. Forbert}
\vskip 14 truept
\centerline{\bigrfont Center for Theoretical Physics and Department 
of Physics}
\vskip 2 truept
\centerline{\bigrfont Texas A\&M University, College Station, TX 77843 USA}
\vskip 1.8 truecm

\centerline{\bf 1.  INTRODUCTION}
\vskip 12 truept

Bose-Einstein condensation is the macroscopic occupation of a single
quantum state. In bulk liquid Helium, most theoretical calculations[1]
agree that, in the limit of zero temperature, roughly 10\% of the Helium
atoms are condensed in the ground state. For recently observed atomic
condensates in $^{87}$Rb [2], $^7$Li [3] and $^{23}$Na [4],
nearly 100\% condensation is expected because of their low density. This
collective occupation essentially magnifies ground state properties by a
factor $N$ equal to the number of condensed atoms. For macroscopic
$N\approx 10^{24}$, this exceedingly large factor can be exploited to
detect minute changes in the condensate ground state. Thus aside from its
intrinsic interest, BEC is potentially a very sensitive probe.

In this work, we will explore changes in the condensate ground state
induced by a ``sizable" impurity. By ``sizable", we mean an impurity
whose size is within a few orders of magnitude of the trap size, and not
necessarily atomic in scale. A sizable impurity will ``drill" a hole in
the condensate wave function  and alter its energy. The question is
whether this microscopic change can be detected macroscopically because
of the Bose-Einstein condensation effect. This is described in the next
section. We discuss the effects of interaction in Section 3 and the
case of two or more impurities in Section 4.

\vskip 24 truept
\centerline{\bf 2. IMPURITY EXPULSION}
\vskip 12 truept

The effect we seek to describe is generic. We will therefore consider
the problem in its simplest conceptual context. We imagine a
non-interacting Bose gas of mass $m$ confined to a spherical cavity (an
infinite square well)  of radius $b$.  We introduce an impurity, which to
first order approximation, can be regarded as  a hard sphere of radius
$a$.  When the impurity is placed inside the cavity, the
wave function of the Bose gas must vanish on the surface of both the
cavity and the hard sphere. Since the cavity only serves to confine the
Bose gas, it may or may not also trap the impurity. Classically, since
there is no interaction between the cavity and the  hard sphere, the
hard sphere is free to be anywhere inside the cavity. However, when the
cavity is filled by a Bose gas, the position of the impurity is dictated
by the ground state energy of the Bose gas. Thus there is a quantum
mechanically induced interaction between the hard sphere and the cavity.
\topinsert
\noindent
\vglue 0.2truein
\hbox{
\vbox{\hsize=6truein
\epsfxsize=3truein
\centerline{\epsffile{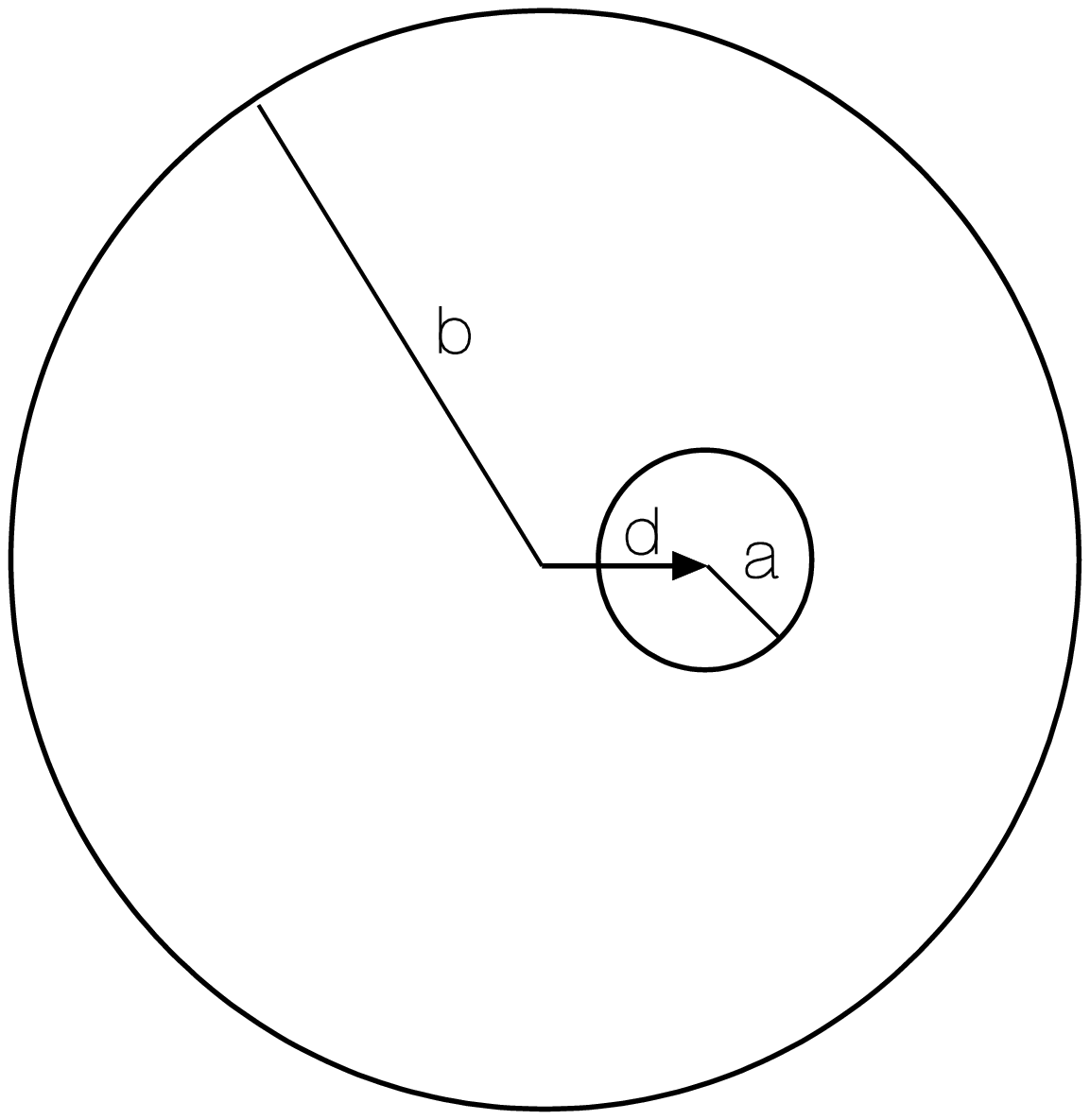}}
}
}
\vglue 0.2truein
\vbox{\hsize=6truein
\baselineskip=12pt
\item{}{\bf Figure 1.}{ The Bose gas is confined in a spherical cavity of
radius $b$ and excluded from an off-center impurity, which is a hard sphere 
of radius $a$.}
}
\vglue 0.3truein
\endinsert

The situation is as shown in Fig. 1, where the hard sphere is displaced by
a distance $d$ from the cavity center. The impurity is assumed to
be sufficiently massive that the Born-Oppenheimer approximation is
adequate. The effective potential for the impurity is then just $V(d)=N
E_0(d)$, where $E_0(d)$ is the ground state energy  of a single particle
confined in a spherical cavity with an off-center hole: 
$$ -{\hbar^2
\over{2m}}{\bf \nabla}^2  \psi({\bf r}) = E(d)\psi({\bf r}),
\qquad\psi({\bf r})=0\quad {\rm at}\quad |{\bf r}|=a\quad {\rm and}
\quad |{\bf r}+{\bf d}|=b.\eqno(1)$$ 
When $d=0$, both the ground state
wave function and the energy are known analytically 
$$\psi_0(r)={1\over
r}\sin\Bigl[{{\pi(r-a)}\over{(b-a)}}\Bigr],\eqno(2)$$
$$E_0={\hbar^2
\over{2m}}{{\pi^2}\over{(b-a)^2}}.\eqno(3)$$ 
When the impurity is
off-center, spherical symmetry is broken and the problem
is non-trivial. Instead of solving the problem exactly, we 
gauge this off-center effect by a variational calculation using 
the following trial function 
$$\phi({\bf r})=\bigl(1-{a\over r}\bigr)(b-|{\bf r}+{\bf d}|).\eqno(4)$$ 
The first factor is the zero energy solution (Laplace's
equation) satisfying the Dirichlet condition at $|{\bf r}|=a$. The
second factor forces the wave function to vanish at the displaced cavity surface.
For $d=0$, the resulting variational energy simply replaces the 
factor $\pi^2$ in (3)
by 10, which is an excellent approximation. For finite $d$, the
resulting energies are as shown in Fig. 2. The angular integration 
is sufficiently cumbersome that we computed the variational energy
by the method of Monte Carlo sampling. The energy is lower as the
impurity moves off center. This result is easy to understand once one
has seen it. The impurity ``drills'' a hole in the cavity's ground state
wave function. It is more costly to drill a hole near the wave function's
maximum (the center) than at its minimum (the edge). Thus a sizable
impurity will be expelled by a Bose condensate.
\topinsert
\noindent
\vglue 0.2truein
\hbox{
\vbox{\hsize=3.0truein
\epsfxsize=3.0truein
\noindent\epsffile{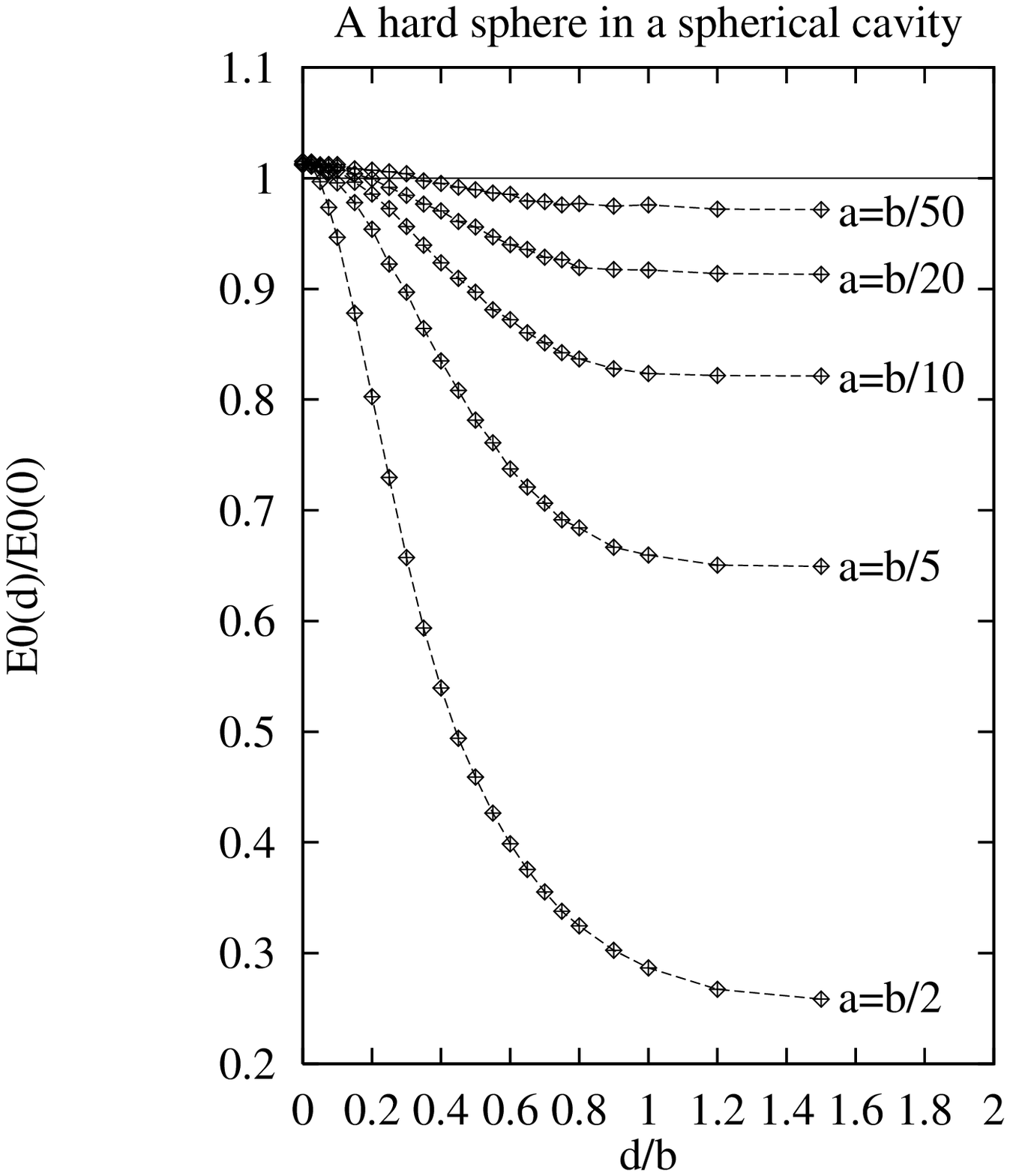}}
\vbox{\hsize=3.0truein
\epsfxsize=3.0truein
\noindent\epsffile{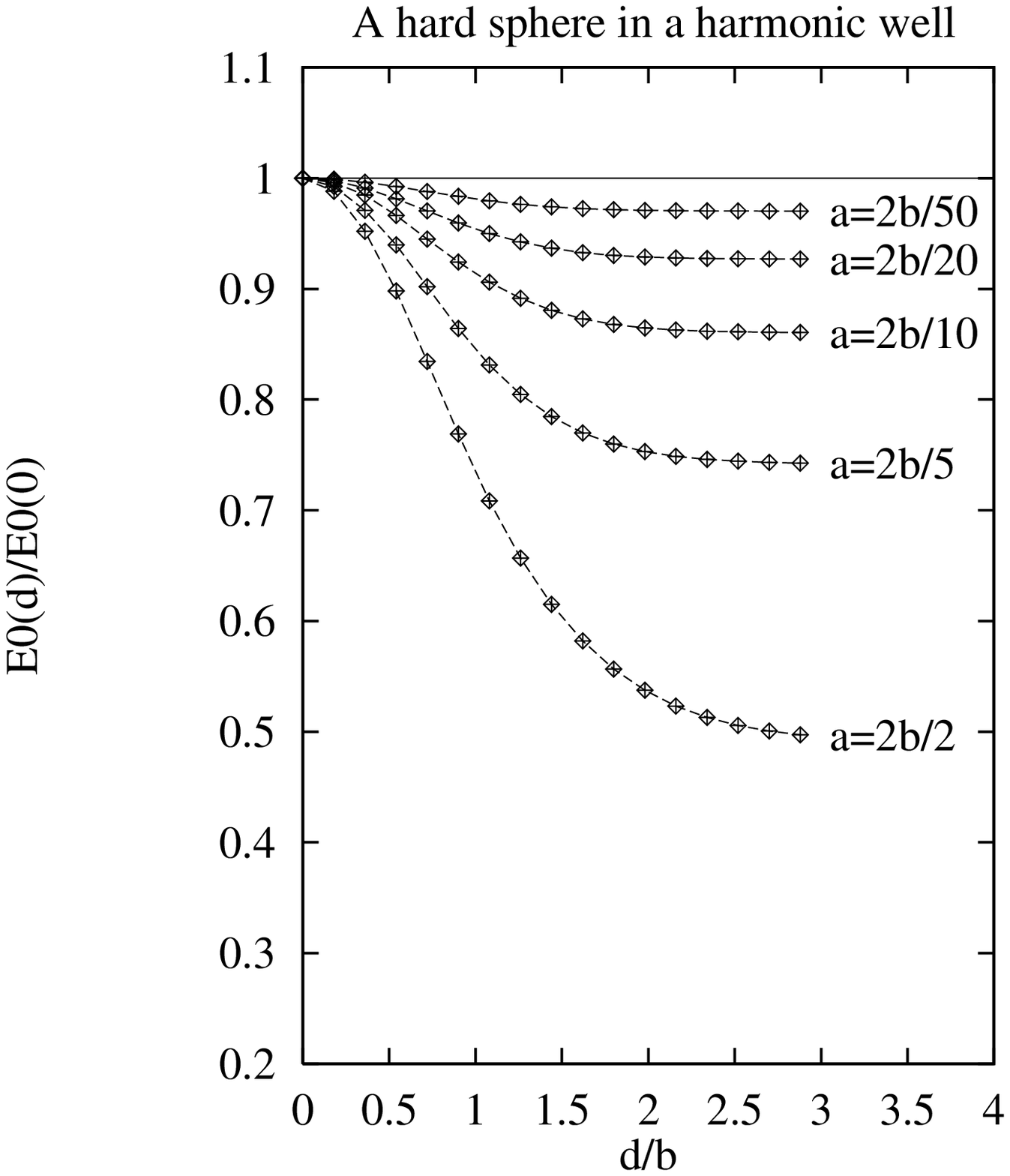}}
}
\vskip 0.1truein
\vbox{\hsize=6truein
\baselineskip=12pt
\item{}{\bf Figure 2.}{ {\it Left}: The ground state energy of a
particle in a spherical cavity of radius $b$ with a displaced hard
sphere of radius $a$. The energy as a function of the displacement for
various impurity sizes is normalized by dividing by the exact on-center
energy. {\it Right}: Similar calculations for the case of an impurity in
a harmonic well. In this case $b$  is the harmonic length and the energy
is only normalized by dividing by the variational on-center energy. In
both cases, the impurity is not assumed to be trapped by the cavity or
the well.}
}
\vglue 0.3truein
\endinsert

Similar results are obtained if one replaces the spherical cavity by a
spherical harmonic oscillator. The size paramenter $b$ can be identified as
the harmonic length via 
 $$\hbar\omega={\hbar^2 \over m}{1\over b^2}.\eqno(5)$$
The harmonic oscillator ground state wave function is then
$$\psi_0({\bf r})=\exp[-{1\over 2}({r\over b})^2].\eqno(6)$$ 
When an impurity is placed in this harmonic well, the ground state 
wave function is again altered. We now do a
variational calculation with the trial function 
$$\phi({\bf
r})=\bigl(1-{a\over r}\bigr) \exp[-{1\over 2}({{|{\bf r}+{\bf d}|}\over
\alpha})^2],\eqno(7)$$ 
where $\alpha$ is a variational parameter. 
The angular integral can be done exactly when we displace the harmonic
oscillator rather than the impurity. The remaining radial integral can then
be computed by numerical quadrature.
The results are
also shown in Fig. 2. In order to compare this with the cavity case, we
must take at least $2b$, and not just $b$, as the equivalent cavity
radius. Since the exact on-center energy in this case has no simple
analytic form, we normalize by dividing by the variational on-center energy.

To get an order-of-magnitude estimate of the expulsion force, consider
the case of lithium atoms with ${{\hbar^2}\over m}\approx 7$ K \AA$^2$ and
$b\approx 3\times 10^4$ \AA. The characteristic energy scale is
$$\hbar\omega={\hbar^2 \over m}{1\over b^2}\approx{7\over{(3\times
10^4)^2}} \approx10^{-8} {\rm K}.$$ 
The variation in energy is over some
fraction of the trap radius. Thus the expulsion force due to each atom is 
$$F
\approx{{10^{-8}{\rm K}}\over{10^4{\rm\AA}}}
\approx{{10^{-8}{\rm K}\, 10^{-13}{\rm J/K}}\over{10^4{\rm\AA}\, 
10^{-10} m/{\rm\AA}}}
\approx 10^{-25}\; {\rm Newton}.$$ 
For $N\approx 10^{24}$ Bose condensed atoms, the expulsion force IS
macroscopic. Currently, the maximum $N$ achieved in any of the atomic
trap experiment is only about $10^6$. Such a small force
may still be detectable if the impurity is sufficiently light.

\vskip 24 truept
\centerline{\bf 3. THE EFFECT OF INTERACTION}
\vskip 12 truept

To gauge the effect of interaction among the Bose condensed
atoms, we do a variational Hartree calculation. For interacting Bosons
confined in a harmonic well, the Hamiltonian is
$$H={\hbar^2 \over {2m}}\sum_i(-\nabla^2_i 
         + {r^2_i\over b^4})
+\sum_{i>j}v({\bf r}_{ij}).\eqno(8)$$
In the low density limit, the two-body potenial can be
replaced by the scattering length approximation
$$v({\bf r}_{ij})\rightarrow 4\pi{\hbar^2\over m}a_{sc}
\delta^3({\bf r}_{ij}).\eqno(9)$$
For a Hartree wave function consisting of a product of
{\it normalized} single particle states $\phi({\bf r}_i)$
$$\Psi({\bf r}_1,{\bf r}_2\dots {\bf r}_n)
=\prod_{i=1}^n\phi({\bf r}_i),\eqno(10)$$
the variational energy is given by
$$
{E_V\over N}={\hbar^2\over {2m}}\Bigl[
\int d^3r\phi({\bf r})(-\nabla^2+{r^2\over b^4})
\phi({\bf r})
   +(N-1)4\pi a_{sc}
\int\! d^3r\phi^2({\bf r})\phi^2({\bf r})\Bigr].
\eqno(11)$$
Minimizing this with respect to $\phi({\bf r})$ yields the
Gross-Pitaevskii equation. To account for the hard sphere impurity, one
must again require $\phi({\bf r})$ to vanish on the impurity's surface.
Instead of solving this problem exactly, we simply take
$$\phi({\bf r})={ 1\over{\sqrt Z}}\bigl(1-{a\over{|{\bf r}+{\bf d}|}}\bigr)
\exp[-{1\over 2}({r\over
\alpha})^2],\eqno(12)$$ 
where $Z$ is the normalization integral, and minimize the 
energy functional(11) with respect to the parameter $\alpha$.
This is in the spirit of using Gaussian trial wave functions to
study the Gross-Pitaevskii equation, as suggested by Baym and Pethick [5]
in the context of BEC.

The interaction is characterized by a strength parameter
$g=N a_{sc}/b$. The left panel of Fig. 3 shows the effect for
positive scattering length. For $g<5$, the results are only slightly
higher than those in the last section. All atomic experiments are far
below the $g=5$ limit. In the extreme case of $g\ge 10$, the picture
changes completely and the impurity is stabilized at the center.
However, when the effective interaction is this strong, the mean-field 
approximation is no longer creditable and one must consider two-body
correlations, as in the case of liquid Helium.
\topinsert
\noindent
\vglue 0.2truein
\hbox{
\vbox{\hsize=3.0truein
\epsfxsize=3.0truein
\noindent\epsffile{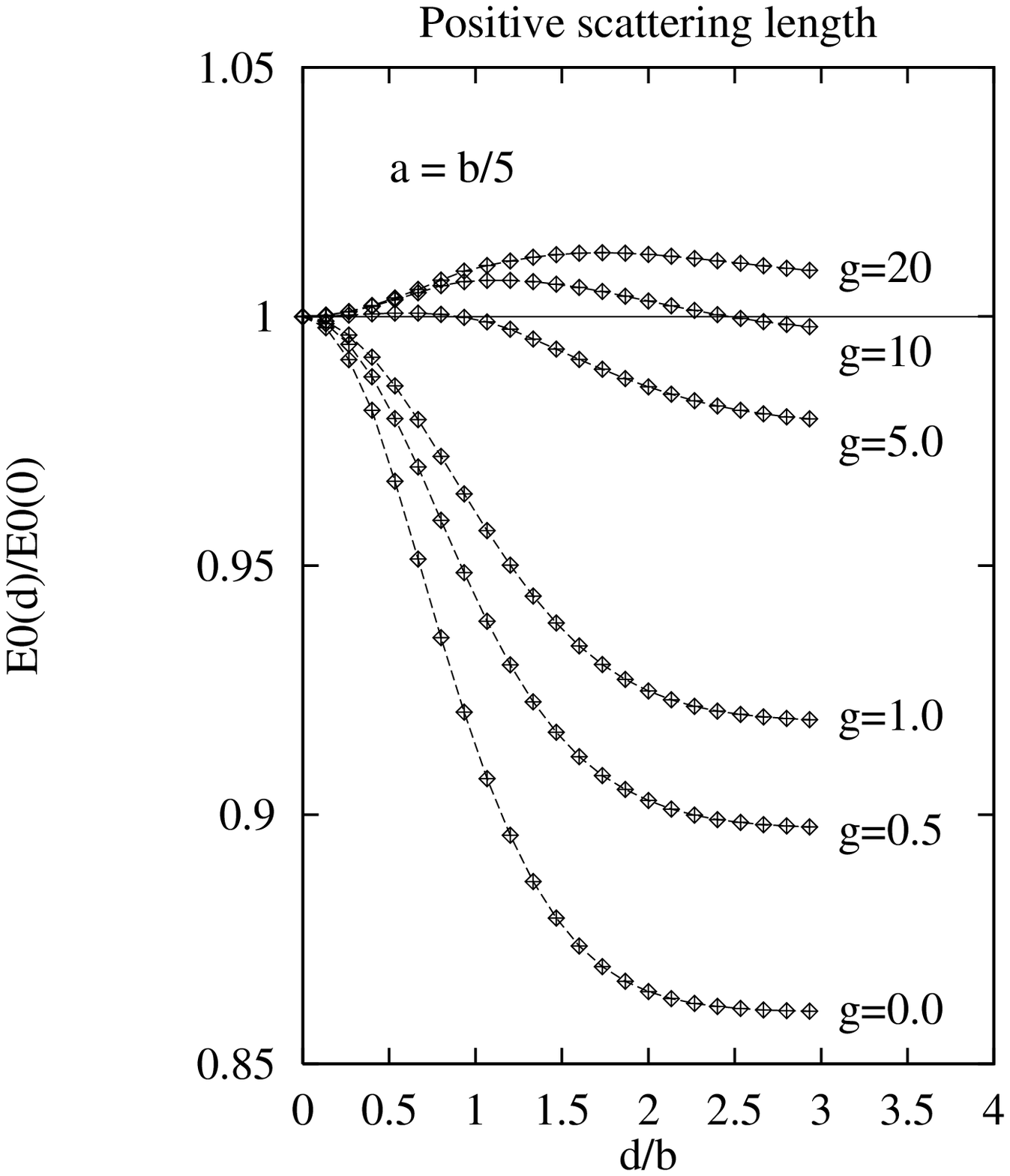}}
\vbox{\hsize=3.0truein
\epsfxsize=3.0truein
\noindent\epsffile{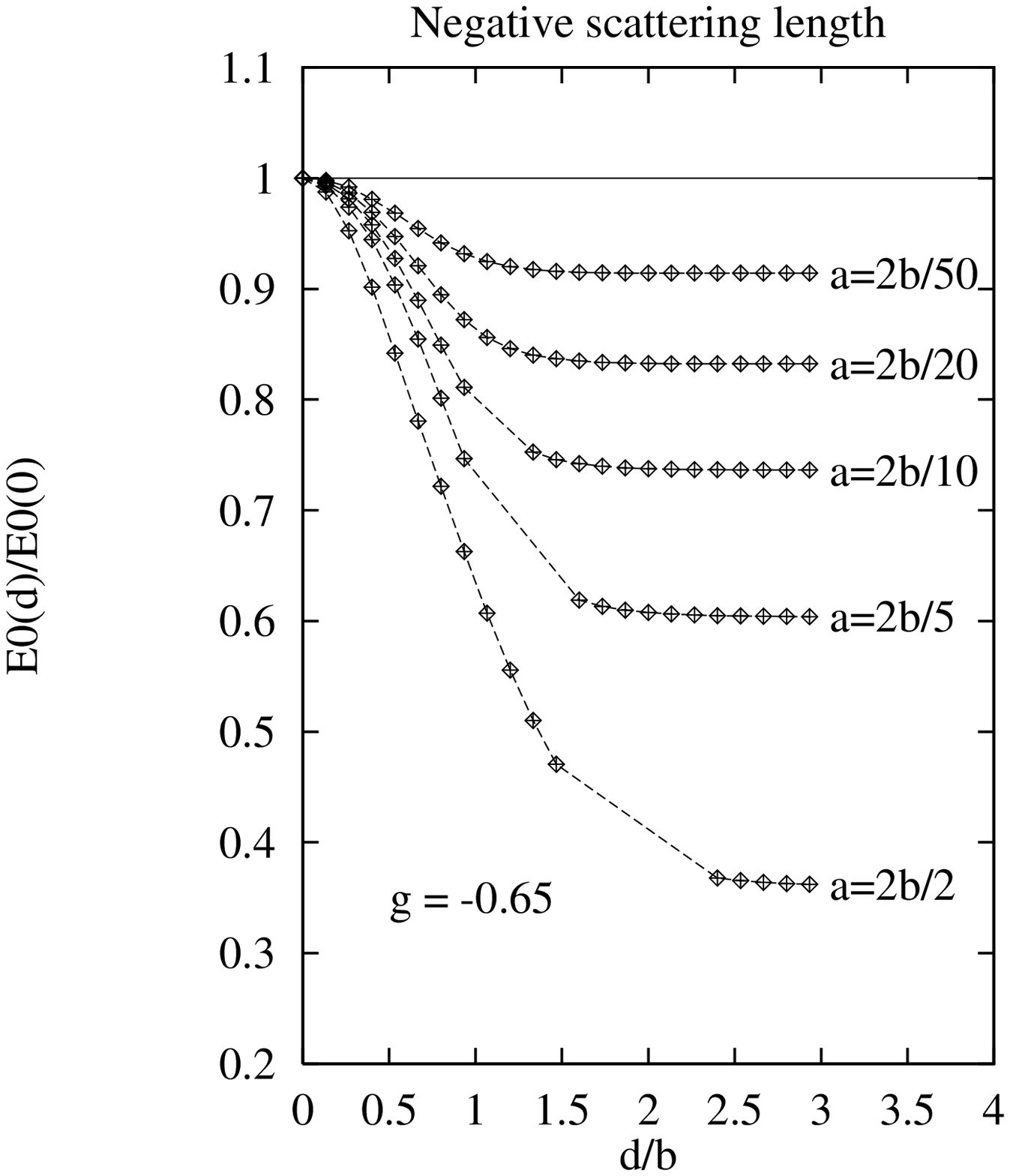}}
}
\vskip 0.1truein
\vbox{\hsize=6truein
\baselineskip=12pt
\item{}{\bf Figure 3.}{ {\it Left}: The change in energy at one
impurity size $a=b/5$ as a function of the interaction
strength $g=N a_{sc}/b$. {\it Right}: The change in energy for 
various impurity sizes at one negative interaction
strength $g=-0.65$. Variationally, the system would collapse at
$g=-0.67$ without any impurity. The straight segments correspond to
impurity locations that would destabilize the condensate. }
}
\vglue 0.1truein
\endinsert

The effect of negative scattering length is shown on the right
panel of Fig. 3. Variationally, as shown by Fetter [6], the condensate 
by itself is unstable
below $g<-0.67$ . When $g$ is close but above this critical value,
the introduction of a sufficiently large impurity will cause the
the condensate to collapse. This is shown by the straight line
segements on Fig. 3. For these locations of the impurity, there are
no energy minima for the variational parameter $\alpha$. When the
impurity is at the center, it increases the single particle energy and there
is no collapse. However, as it moves off center, the energy is lowered and
the condensate collapses. This is also understandable from the opposite 
direction; when $g$ is close
to the critical value, as the impurity enters the condensate, it increases
the condensate density without substantially increasing the single particle energy.
It pushes $g$ over the critical value and the condensate collapses.

\vskip 24 truept
\centerline{\bf 4. TWO OR MORE IMPURITIES}
\vskip 12 truept

For the case of $n$ impurities, each having a different
radius $a_i$ and located at ${\bf d}_i$, one can simply generalize
the single particle trial function (12) to 
$$\phi({\bf r})={ 1\over{\sqrt Z}}\prod_{i=1}^nf_i({\bf r})
\exp[-{1\over 2}({r\over
\alpha})^2],\eqno(13)$$ 
with
$$
f_i({\bf r})=\bigl(1-{a_i\over{|{\bf r}+{\bf d}_i|}}\bigr).\eqno(14)
$$
The resulting expression for the variational energy
has a simple form
$$
{E_V\over N}={\hbar^2\over {m}}
\Bigl[ {3\over 2}{1\over\alpha^2}
   +{1\over 2}\Bigl\langle 
   \sum_{i=1}^n{\bf g}_i^2 
   -\Bigl(\sum_{i=1}^n{\bf g}_i-{ {\bf r}\over\alpha^2}\bigr)^2
   +{r^2\over b^4}\Bigr\rangle 
\Bigr],
\eqno(15)$$
where
$${\bf g}_i={{\nabla f_i}\over{f_i}},\eqno(16)$$
and the expectation value is with respect to the trial function (13).
The first two terms in the expectation value involving ${\bf g}_i$
may be interpreted as the kinetic energy of each impurity and their
collective interaction with the harmonic well. With more than
one impurity, there is no way to do the angular integration exactly.
We evaluated the expectation value in (15) by the Monte Carlo method. 

\topinsert
\noindent
\vglue 0.2truein
\hbox{
\vbox{\hsize=2.5truein
\epsfxsize=2.5truein
\noindent\epsffile{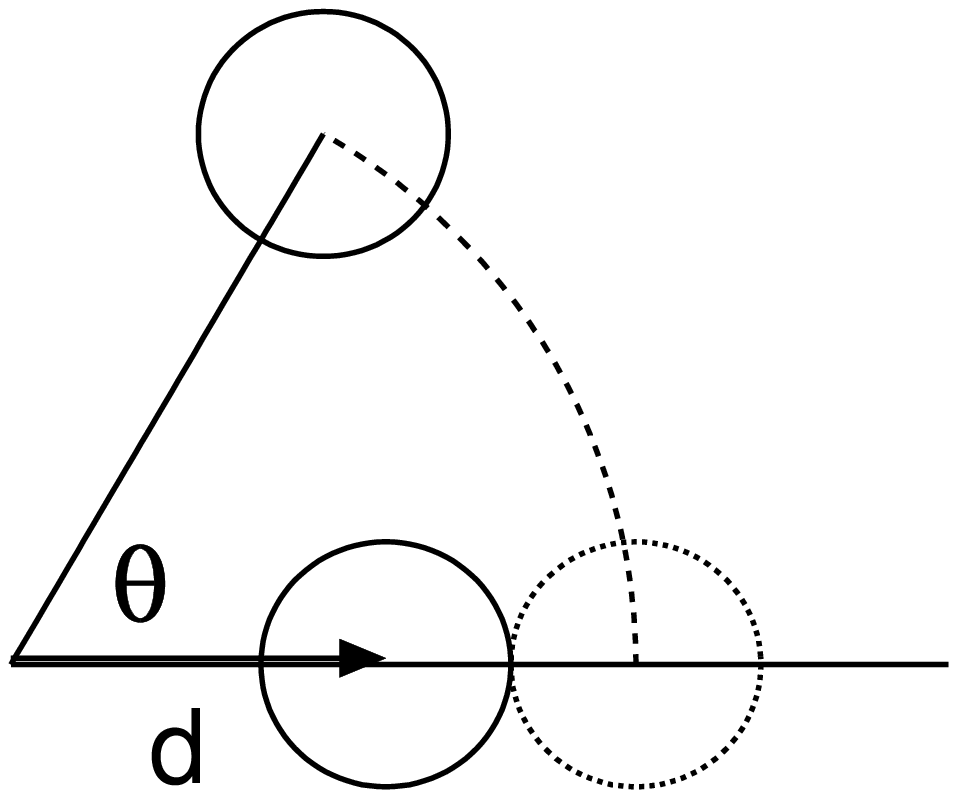}}
\vbox{\hsize=3.5truein
\epsfxsize=3.5truein
\noindent\epsffile{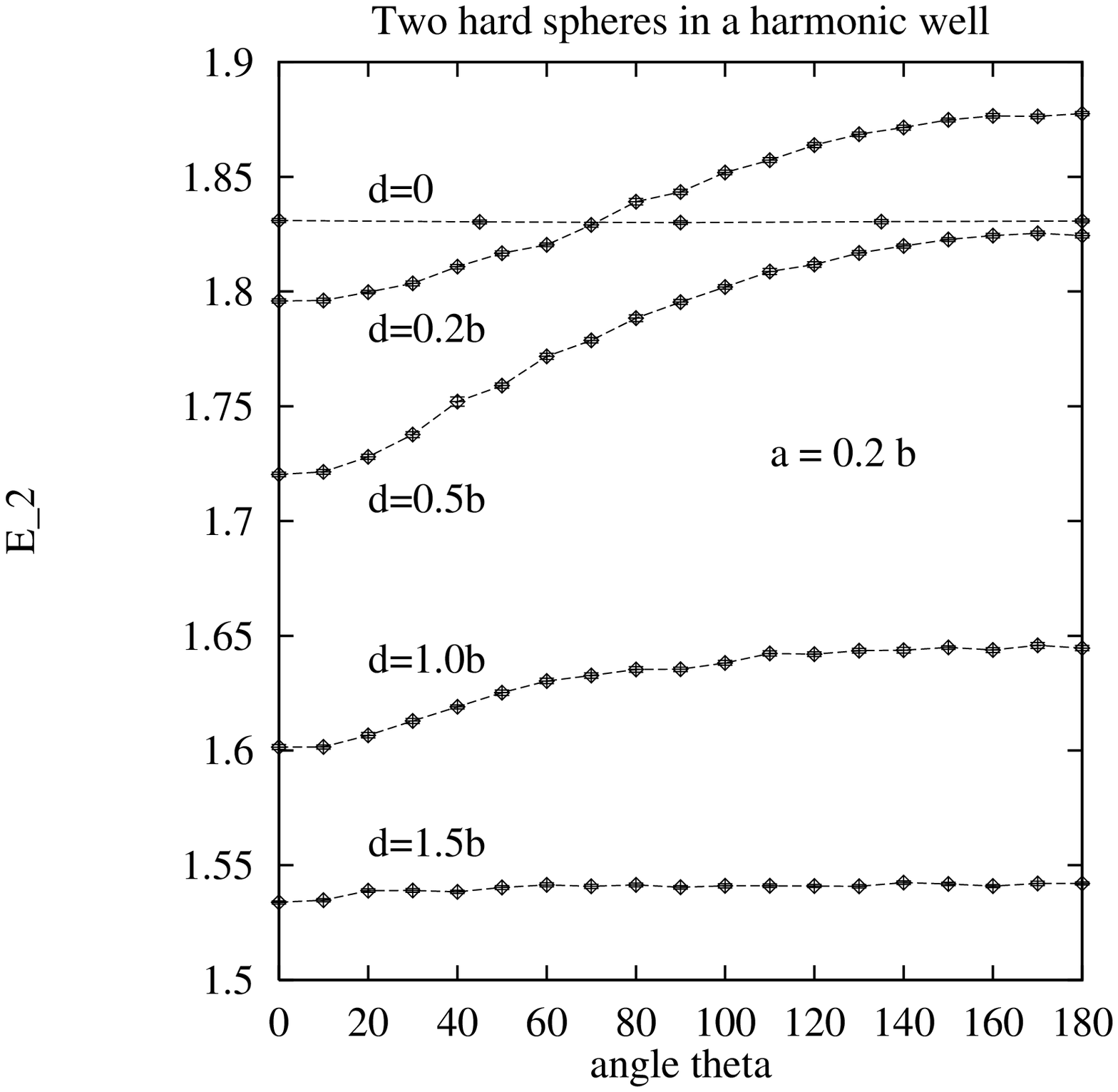}}
}
\vskip 0.1truein
\vbox{\hsize=6truein
\baselineskip=12pt
\item{}{\bf Figure 4.}{ {\it Left}: Two impurities maintaining the
same distance from the harmonic well center while rotating through
an angular separation $\theta$.
 {\it Right}: The energy for two impurities, each of size $a=0.2 b$,
as a function of angular
separation at various distances from the well center. }
}
\vglue 0.3truein
\endinsert
For $n=2$, the
resulting energy $E_2=E_V/\hbar\omega$ is shown in Fig. 4.
In order to disentangle the two-impurity interaction energy from the 
effect of off-center energy dependence, we separate the two impurities
by keeping each at equal distance from the trap center. The configuration
used is as shown on the left panel of Fig. 4. The resulting energy
on the right panel clearly shows that there is an effective attraction
between the two impurities. This is induced by the Bose condensate.
It is less costly to ``drill'' a slightly larger hole at one place than
to ``drill" two separated holes. One can therefore infer that multiple
impurities tend to clump together and will be expelled
from the trap center.

The effect of interaction can again be assessed by including a 
Gross-Pitaevskii type mean-field interaction. For the present
qualitative discussion, we have not bothered to include this correction.

\vskip 24 truept
\centerline{\bf 5. CONCLUSIONS }
\vskip 12 truept

    In this work, we have considered the possible use of BEC as
a sensitive probe for detecting microscopic changes in the
condensate ground state. Our variational studies suggest that

\vskip 12 truept
\itemitem{a)} A hard-sphere-like impurity will be expelled
from the center of a condensate. For a light but sizable impurity, such
an expulsion maybe macroscopically observable.

\itemitem{b)} A sizable, hard-sphere-like impurity will accelerate
the collapse of a condensate with negative scattering length.

\itemitem{c)} A Bose condensate induces an effective attraction among
hard-sphere-like impurities. This may have interesting implications
for induced dimerization or clusterization of weakly interacting
impurities, such as $^3$He.
\vskip 12 truept

Work is currently in progress to seek exact solutions for problems that
have only been solved variationally in this work.

\vskip 24 truept
\centerline{\bf ACKNOWLEDGMENTS}
\vskip 12 truept

This research was funded, in part, by the U. S. National Science Foundation 
grants PHY95-12428 and DMR-9509743 (to SAC). The idea of impurity expulsion
was evolved from considerations of impurity delocalization in Helium
droplets[7-9]. The latter was first suggested by my colleague and collaborator 
E. Krotscheck.

\vskip 28 truept
\centerline{\bf REFERENCES}
\vskip 12 truept

\item{[1]} R. M. Panoff and P. A. Whitlock, in {\it Momentum Distribution},
           edited by R. N. Silver and P. E. Sokol, Plenum Publishing, 1989.
\item{[2]} M. H. Anderson {\it et al.}, Science {\bf 269}, 198 (1995)
\item{[3]} C. C. Bradley {\it et al.}, Phys. Rev. Lett. {\bf 75}, 1687 (1995);
           {\bf 78}, 985 (1997)
\item{[4]} K. B. Davis {\it et al.}, Phys. Rev. Lett. {\bf 75}, 3969 (1995)
\item{[5]} G. Baym and C. Pethick, Phys. Rev. Lett. {\bf 76}, 2477 (1996)
\item{[6]} A. L. Fetter, ``Ground State and Excited States of a
          confined Bose Gas," cond-mat/9510037.
\item{[7]}E. Krotscheck and S.~A. Chin,
		Chem. Phys. Lett. {\bf 227}, 143 (1994).
\item{[8]}S. A. Chin and E. Krotscheck,
		Phys. Rev. B52, 10405 (1995)
\item{[9]} S. A. Chin and E. Krotscheck,
		{\it Recent Progress in Many-Body Theories,\/} Vol. 4,
                P.85, edited by E. Schachinger {\it et al.}, Plenum Press,
                1995.
\vfill

\end